\pdfoutput=1
\documentclass[12pt,a4paper]{article}

\usepackage{ifthen} 
\newboolean{pdflatex}
\setboolean{pdflatex}{true} 

\newboolean{articletitles}
\setboolean{articletitles}{true} 

\newboolean{uprightparticles}
\setboolean{uprightparticles}{false} 

\def\paperauthors{LHCb collaboration} 
\def\paperasciititle{Observation of the suppressed Lb -> D0pK- decay with D -> K+pi- and measurement of its CP asymmetry} 
\def\papertitle{Observation of the suppressed $\Lb\to D\proton\Km$ decay with $D\to\Kp\pim$ and measurement of its \CP asymmetry} 
\def\paperkeywords{{High Energy Physics}, {LHCb}} 
\def\papercopyright{\the\year\ CERN for the benefit of the LHCb collaboration} 
\def\paperlicence{CC BY 4.0 licence}
\def\paperlicenceurl{https://creativecommons.org/licenses/by/4.0/}

\usepackage[top=1in, bottom=1.25in, left=1in, right=1in]{geometry}

\usepackage{microtype}
\usepackage{lineno}  
\usepackage{xspace} 
\usepackage{caption} 

\usepackage{graphicx}  
\usepackage{color}
\usepackage{colortbl}
\graphicspath{{./figs/}} 

\usepackage{amsmath} 
\usepackage{amssymb}
\usepackage{amsfonts}
\usepackage{upgreek} 

\newcommand*\patchAmsMathEnvironmentForLineno[1]{%
\expandafter\let\csname old#1\expandafter\endcsname\csname #1\endcsname
\expandafter\let\csname oldend#1\expandafter\endcsname\csname
end#1\endcsname
 \renewenvironment{#1}%
   {\linenomath\csname old#1\endcsname}%
   {\csname oldend#1\endcsname\endlinenomath}%
}
\newcommand*\patchBothAmsMathEnvironmentsForLineno[1]{%
  \patchAmsMathEnvironmentForLineno{#1}%
  \patchAmsMathEnvironmentForLineno{#1*}%
}
\AtBeginDocument{%
\patchBothAmsMathEnvironmentsForLineno{equation}%
\patchBothAmsMathEnvironmentsForLineno{align}%
\patchBothAmsMathEnvironmentsForLineno{flalign}%
\patchBothAmsMathEnvironmentsForLineno{alignat}%
\patchBothAmsMathEnvironmentsForLineno{gather}%
\patchBothAmsMathEnvironmentsForLineno{multline}%
\patchBothAmsMathEnvironmentsForLineno{eqnarray}%
}

\usepackage{hyperxmp}

\usepackage[pdftex,
            pdfauthor={\paperauthors},
            pdftitle={\paperasciititle},
            pdfkeywords={\paperkeywords},
            pdfcopyright={Copyright (C) \papercopyright},
            pdflicenseurl={\paperlicenceurl}]{hyperref}

\usepackage[colorinlistoftodos,textsize=scriptsize]{todonotes}

\usepackage[bottom,flushmargin,hang,multiple]{footmisc}

\usepackage[all]{hypcap} 

\usepackage{xspace} 
\usepackage{upgreek}

\def\lhcb   {\mbox{LHCb}\xspace}

\def\MagUp {\mbox{\em Mag\kern -0.05em Up}\xspace}

\ifthenelse{\boolean{uprightparticles}}%
{

 \def\Ppi         {\ensuremath{\uppi}\xspace}

 \def\PDelta      {\ensuremath{\Delta}\xspace}                 
 \def\PXi         {\ensuremath{\Xi}\xspace}                 
 \def\PLambda     {\ensuremath{\Lambda}\xspace}                 
 \def\PSigma      {\ensuremath{\Sigma}\xspace}                 
 \def\POmega      {\ensuremath{\Omega}\xspace}                 
 \def\PUpsilon    {\ensuremath{\Upsilon}\xspace}

 \def\PB      {\ensuremath{\mathrm{B}}\xspace}                 
                  
 \def\PD      {\ensuremath{\mathrm{D}}\xspace}

 \def\PK      {\ensuremath{\mathrm{K}}\xspace}

 \def\Pb      {\ensuremath{\mathrm{b}}\xspace}                 
 \def\Pc      {\ensuremath{\mathrm{c}}\xspace}

 \def\Pi      {\ensuremath{\mathrm{i}}\xspace}

 \def\Pp      {\ensuremath{\mathrm{p}}\xspace}

 \def\Ps      {\ensuremath{\mathrm{s}}\xspace}

 \def\thebaroffset{0.0em}
}
{

 \def\Ppi         {\ensuremath{\pi}\xspace}

 \mathchardef\PDelta="7101
 \mathchardef\PXi="7104
 \mathchardef\PLambda="7103
 \mathchardef\PSigma="7106
 \mathchardef\POmega="710A
 \mathchardef\PUpsilon="7107
                  
 \def\PB      {\ensuremath{B}\xspace}                 
                  
 \def\PD      {\ensuremath{D}\xspace}

 \def\PK      {\ensuremath{K}\xspace}

 \def\Pb      {\ensuremath{b}\xspace}                 
 \def\Pc      {\ensuremath{c}\xspace}

 \def\Pi      {\ensuremath{i}\xspace}

 \def\Pp      {\ensuremath{p}\xspace}

 \def\Ps      {\ensuremath{s}\xspace}

 \def\thebaroffset{0.18em}
}
\newcommand{\offsetoverline}[2][\thebaroffset]{\kern #1\overline{\kern -#1 #2}}%

\makeatletter
\ifcase \@ptsize \relax
  \newcommand{\miniscule}{\@setfontsize\miniscule{4}{5}}
\or
  \newcommand{\miniscule}{\@setfontsize\miniscule{5}{6}}
\or
  \newcommand{\miniscule}{\@setfontsize\miniscule{5}{6}}
\fi
\makeatother

\DeclareRobustCommand{\optbar}[1]{\shortstack{{\miniscule (\rule[.5ex]{1.25em}{.18mm})}
  \\ [-.7ex] $#1$}}

\def\squark    {{\ensuremath{\Ps}}\xspace}

\def\cquark    {{\ensuremath{\Pc}}\xspace}

\def\bquark    {{\ensuremath{\Pb}}\xspace}

\def\pion   {{\ensuremath{\Ppi}}\xspace}
\def\piz    {{\ensuremath{\pion^0}}\xspace}
\def\pip    {{\ensuremath{\pion^+}}\xspace}
\def\pim    {{\ensuremath{\pion^-}}\xspace}

\def\pimp   {{\ensuremath{\pion^\mp}}\xspace}

\def\kaon    {{\ensuremath{\PK}}\xspace}

\def\KorKbar {\kern \thebaroffset\optbar{\kern -\thebaroffset \PK}{}\xspace}

\def\Kp      {{\ensuremath{\kaon^+}}\xspace}
\def\Km      {{\ensuremath{\kaon^-}}\xspace}
\def\Kpm     {{\ensuremath{\kaon^\pm}}\xspace}

\def\Dbar    {{\ensuremath{\offsetoverline{\PD}}}\xspace}
\def\D       {{\ensuremath{\PD}}\xspace}

\def\DorDbar {\kern \thebaroffset\optbar{\kern -\thebaroffset \PD}\xspace}
\def\Dz      {{\ensuremath{\D^0}}\xspace}
\def\Dzb     {{\ensuremath{\Dbar{}^0}}\xspace}
\def\Dp      {{\ensuremath{\D^+}}\xspace}
\def\Dm      {{\ensuremath{\D^-}}\xspace}

\def\DpDm    {\ensuremath{\Dp {\kern -0.16em \Dm}}\xspace}
\def\Dstar   {{\ensuremath{\D^*}}\xspace}

\def\Dstarz  {{\ensuremath{\D^{*0}}}\xspace}

\def\B       {{\ensuremath{\PB}}\xspace}

\def\BorBbar {\kern \thebaroffset\optbar{\kern -\thebaroffset \PB}\xspace}

\def\Bd      {{\ensuremath{\B^0}}\xspace}

\def\BdorBdbar {\kern \thebaroffset\optbar{\kern -\thebaroffset \Bd}\xspace}

\def\Bs      {{\ensuremath{\B^0_\squark}}\xspace}

\def\BsorBsbar {\kern \thebaroffset\optbar{\kern -\thebaroffset \Bs}\xspace}

\def\Y#1S{\ensuremath{\PUpsilon{(#1S)}}\xspace}

\def\proton      {{\ensuremath{\Pp}}\xspace}
\def\antiproton  {{\ensuremath{\overline \proton}}\xspace}

\def\Lz          {{\ensuremath{\PLambda}}\xspace}
\def\Lbar        {{\ensuremath{\offsetoverline{\PLambda}}}\xspace}
\def\LorLbar     {\kern \thebaroffset\optbar{\kern -\thebaroffset \PLambda}\xspace}
\def\Lambdares   {{\ensuremath{\PLambda}}\xspace}

\def\Xires       {{\ensuremath{\PXi}}\xspace}

\def\Lc          {{\ensuremath{\Lz^+_\cquark}}\xspace}

\def\Lb           {{\ensuremath{\Lz^0_\bquark}}\xspace}
\def\Lbbar        {{\ensuremath{\Lbar{}^0_\bquark}}\xspace}

\def\Xib          {{\ensuremath{\Xires_\bquark}}\xspace}
\def\Xibz         {{\ensuremath{\Xires^0_\bquark}}\xspace}

\def\to                 {\ensuremath{\rightarrow}\xspace}

\def\CP                {{\ensuremath{C\!P}}\xspace}

\def\AT#1     {\ensuremath{A_{\mathrm{T}}^{#1}}\xspace}           

\def\C#1      {\ensuremath{\mathcal{C}_{#1}}\xspace}                       
\def\Cp#1     {\ensuremath{\mathcal{C}_{#1}^{'}}\xspace}                    
\def\Ceff#1   {\ensuremath{\mathcal{C}_{#1}^{\mathrm{(eff)}}}\xspace}        
\def\Cpeff#1  {\ensuremath{\mathcal{C}_{#1}^{'\mathrm{(eff)}}}\xspace}       
\def\Ope#1    {\ensuremath{\mathcal{O}_{#1}}\xspace}                       
\def\Opep#1   {\ensuremath{\mathcal{O}_{#1}^{'}}\xspace}                    

\newcommand{\nospaceunit}[1]{\ensuremath{\text{#1}}}       
\newcommand{\aunit}[1]{\ensuremath{\text{\,#1}}}       

\newcommand{\tev}{\aunit{Te\kern -0.1em V}\xspace}
\newcommand{\gev}{\aunit{Ge\kern -0.1em V}\xspace}
\newcommand{\mev}{\aunit{Me\kern -0.1em V}\xspace}
\newcommand{\kev}{\aunit{ke\kern -0.1em V}\xspace}
\newcommand{\ev}{\aunit{e\kern -0.1em V}\xspace}
 
\newcommand{\mevc}{\ensuremath{\aunit{Me\kern -0.1em V\!/}c}\xspace}
\newcommand{\gevc}{\ensuremath{\aunit{Ge\kern -0.1em V\!/}c}\xspace}
\newcommand{\mevcc}{\ensuremath{\aunit{Me\kern -0.1em V\!/}c^2}\xspace}
\newcommand{\gevcc}{\ensuremath{\aunit{Ge\kern -0.1em V\!/}c^2}\xspace}
\newcommand{\gevgevcccc}{\ensuremath{\gev^2\!/c^4}\xspace} 

\def\mum  {\ensuremath{\,\upmu\nospaceunit{m}}\xspace}

\def\fb   {\ensuremath{\aunit{fb}}\xspace}
\def\invfb   {\ensuremath{\fb^{-1}}\xspace}

\newcommand{\chisq}{\ensuremath{\chi^2}\xspace}

\newcommand{\chisqip}{\ensuremath{\chi^2_{\text{IP}}}\xspace}

\def\gsim{{~\raise.15em\hbox{$>$}\kern-.85em
          \lower.35em\hbox{$\sim$}~}\xspace}
\def\lsim{{~\raise.15em\hbox{$<$}\kern-.85em
          \lower.35em\hbox{$\sim$}~}\xspace}

\def\sPlot{\mbox{\em sPlot}\xspace}

\def\pt         {\ensuremath{p_{\mathrm{T}}}\xspace}

\def\ptot       {\ensuremath{p}\xspace}

\def\evtgen     {\mbox{\textsc{EvtGen}}\xspace}

\def\geant      {\mbox{\textsc{Geant4}}\xspace}

\def\photos     {\mbox{\textsc{Photos}}\xspace}

\def\pythia     {\mbox{\textsc{Pythia}}\xspace}

\def\tell1  {TELL1\xspace}
\def\ukl1   {UKL1\xspace}

\usepackage{cite} 
\usepackage{mciteplus}

\usepackage{longtable} 

\usepackage{float}
\usepackage{grffile}
\newcommand{\lbtodpk}{\mbox{\ensuremath{\Lb \to \D \proton \Km}}\xspace}
\newcommand{\lbtodpkfav}{\mbox{\ensuremath{\Lb \to [\Km\pip]_{D} \proton \Km}}\xspace}
\newcommand{\lbtodpksup}{\mbox{\ensuremath{\Lb \to [\Kp\pim]_{D} \proton \Km}}\xspace}
\newcommand{\lbtodpksupbar}{\mbox{\ensuremath{\Lbbar \to [\Km\pip]_{D} \antiproton \Kp}}\xspace}
\usepackage[section]{placeins}
\newcommand{\FAV}{favoured\xspace}
\newcommand{\SUP}{suppressed\xspace}
\newcommand{\anaR}{\ensuremath{7.1}\xspace}
\newcommand{\anaRstat}{\ensuremath{0.8}\xspace}

\newcommand{\anaA}{\ensuremath{0.12}\xspace}
\newcommand{\anaAstat}{\ensuremath{0.09}\xspace}

\newcommand{\anaRlowmpk}{\ensuremath{8.6}\xspace}
\newcommand{\anaRlowmpkstat}{\ensuremath{1.5}\xspace}
\newcommand{\anaAlowmpk}{\ensuremath{0.01}\xspace}
\newcommand{\anaAlowmpkstat}{\ensuremath{0.16}\xspace}

\newcommand{\anaRsysu}{\ensuremath{0.4}\xspace}
\newcommand{\anaRsysd}{\ensuremath{0.3}\xspace}
\newcommand{\anaAsysu}{\ensuremath{0.02}\xspace}
\newcommand{\anaAsysd}{\ensuremath{0.03}\xspace}

\newcommand{\anaRlowmpksysu}{\ensuremath{0.4}\xspace}
\newcommand{\anaRlowmpksysd}{\ensuremath{0.3}\xspace}
\newcommand{\anaAlowmpksysu}{\ensuremath{0.03}\xspace}
\newcommand{\anaAlowmpksysd}{\ensuremath{0.02}\xspace}
\newcommand{\Rmeas}{\ensuremath{\anaR \pm \anaRstat \,(\text{stat.}) {}^{+\anaRsysu}_{-\anaRsysd} \,(\text{syst.})}\xspace}
\newcommand{\Ameas}{\ensuremath{\anaA \pm \anaAstat \,(\text{stat.}) {}^{+\anaAsysu}_{-\anaAsysd} \,(\text{syst.})}\xspace}
\newcommand{\Rlowmpkmeas}{\ensuremath{\anaRlowmpk \pm \anaRlowmpkstat \,(\text{stat.}) {}^{+\anaRlowmpksysu}_{-\anaRlowmpksysd} \,(\text{syst.})}\xspace}
\newcommand{\Alowmpkmeas}{\ensuremath{\anaAlowmpk \pm \anaAlowmpkstat \,(\text{stat.}) {}^{+\anaAlowmpksysu}_{-\anaAlowmpksysd} \,(\text{syst.})}\xspace}
\begin{document}

\renewcommand{\thefootnote}{\fnsymbol{footnote}}
\setcounter{footnote}{1}


\begin{titlepage}
\pagenumbering{roman}

\vspace*{-1.5cm}
\centerline{\large EUROPEAN ORGANIZATION FOR NUCLEAR RESEARCH (CERN)}
\vspace*{1.5cm}
\noindent
\begin{tabular*}{\linewidth}{lc@{\extracolsep{\fill}}r@{\extracolsep{0pt}}}
\ifthenelse{\boolean{pdflatex}}
{\vspace*{-1.5cm}\mbox{\!\!\!\includegraphics[width=.14\textwidth]{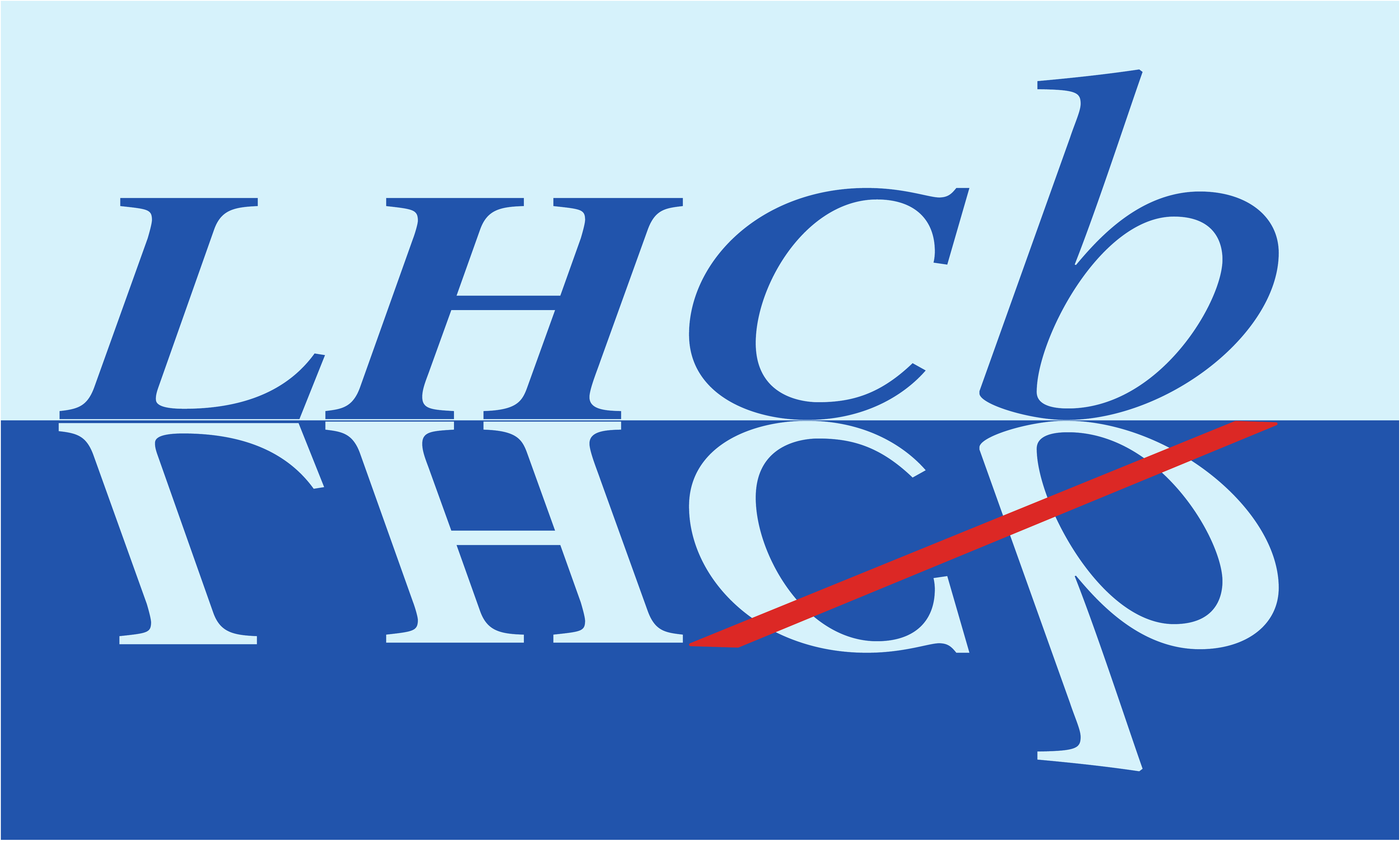}} & &}%
{\vspace*{-1.2cm}\mbox{\!\!\!\includegraphics[width=.12\textwidth]{figs/lhcb-logo.eps}} & &}%
\\
 & & CERN-EP-2021-179 \\  
 & & LHCb-PAPER-2021-027 \\  
 & & 7 January 2022 \\ 
 & & \\
\end{tabular*}

\vspace*{4.0cm}

{\normalfont\bfseries\boldmath\huge
\begin{center}
  \papertitle 
\end{center}
}

\vspace*{2.0cm}

\begin{center}
\paperauthors\footnote{Authors are listed at the end of this paper.}
\end{center}

\vspace{\fill}

\begin{abstract}
  \noindent
A study of \Lb baryon decays to the $Dp\Km$ final state is presented based on a proton-proton collision data sample corresponding to an integrated luminosity of $9\invfb$ collected with the LHCb detector. Two \Lb decays are considered, $\Lb\to D\proton\Km$ with $D\to\Km\pip$ and $D\to\Kp\pim$, where $D$ represents a superposition of \Dz and \Dzb states. The latter process is expected to be suppressed relative to the former, and is observed for the first time. The ratio of branching fractions of the two decays is measured, and the \CP asymmetry of the suppressed mode, which is sensitive to the CKM angle $\gamma$, is also reported.
\end{abstract}

\vspace*{2.0cm}

\begin{center}
  Published in Phys.~Rev.~D104 (2021) 112008
\end{center}

\vspace{\fill}

{\footnotesize 
\centerline{\copyright~\papercopyright. \href{\paperlicenceurl}{\paperlicence}.}}
\vspace*{2mm}

\end{titlepage}


\newpage
\setcounter{page}{2}
\mbox{~}


\renewcommand{\thefootnote}{\arabic{footnote}}
\setcounter{footnote}{0}

\cleardoublepage


\pagestyle{plain} 
\setcounter{page}{1}
\pagenumbering{arabic}


\section{Introduction}
\label{sec:Introduction}

Few studies of beauty-baryon decays to final states involving a single open-charm meson exist, but they are nonetheless promising for measurements of \CP violation~\cite{Dunietz:1992ti, Fayyazuddin:1998aa, Giri:2001ju}. A measurement of a set of branching fraction ratios of \Lb and \Xib decays to final states including a $D$ meson yielded the first observation of the singly-Cabibbo-suppressed \lbtodpkfav decay~\cite{LHCb-PAPER-2013-056}, where $D$ represents a \Dz or \Dzb meson.\footnote{Charge conjugation is implied throughout this document except in reference to \CP asymmetries or comparisons of \Lb and \Lbbar samples.} This was followed by a study of the resonant structure of $\Lb\to\Dz p\pim$ decays~\cite{LHCb-PAPER-2016-061}. This paper reports the results of a study of \lbtodpk decays with the objectives of observing for the first time the $\Lb\to D\proton\Km$ decay with $D\to\Kp\pim$, denoted as \lbtodpksup 
and measuring its 
\CP asymmetry. This decay is
expected to be suppressed relative to the \lbtodpkfav decay. An estimate of the ratio of branching fractions between the
favoured and suppressed modes is obtained by considering the relevant \mbox{Cabibbo-Kobayashi-Maskawa} (CKM) matrix elements~\cite{PDG2020}
\begin{equation}\label{eq_Rpred}
R \approx \left|\frac{V_{cb} V^{*}_{us}}{V_{ub} V^{*}_{cs}}\right|^2 = 6.0.
\end{equation}
The \lbtodpkfav (\lbtodpksup) decay with same (opposite) sign kaons are referred to as the favoured (suppressed) decay throughout this paper. The suppressed decay is of particular interest since its decay amplitude receives contributions from $b\to c$ and $b\to u$ amplitudes of similar magnitude, given the CKM suppression between the two $D$ decays. The interference between these two amplitudes, which depends upon the CKM angle $\gamma$, is expected to be large~\cite{Atwood:1996ci,Atwood:2000ck}, but the different strong phases associated with the various configurations of polarisation states for the \Lb, proton, and intermediate resonances complicate determination of $\gamma$.

The analysis is based on proton-proton ($pp$) collision data collected with the LHCb detector at \mbox{$\sqrt{s}=$ 7, 8, and
13\tev}, corresponding to a total integrated luminosity of 9\invfb. The suppressed \lbtodpksup decay is observed for the first time. In addition the ratio of branching fractions of the \FAV and \SUP decays, $R$, and the \CP asymmetry in the \SUP mode, $A$, which is expected to be sensitive to the CKM angle $\gamma$, are measured where
	\begin{align}
	R &= \frac{{\cal B}(\lbtodpkfav)}{{\cal B}(\lbtodpksup)},\nonumber
	\end{align}
including both flavours, and
    \begin{align}
	A &= \frac{{\cal B}(\Lb \to [\Kp\pim]_{\D} \proton \Km) - {\cal B}(\Lbbar \to [\Km\pip]_{\D} \antiproton \Kp)}{{\cal B}(\Lb \to [\Kp\pim]_{\D} \proton \Km) + {\cal B}(\Lbbar \to [\Km\pip]_{\D} \antiproton \Kp)}.
	\end{align}
Sensitivity to \CP-violation requires interference between amplitudes involving intermediate \Dz and \Dzb mesons. This interference is anticipated to be amplified in regions of the phase space involving $\Lb \to \D X$ contributions, where $X$ labels excited $\Lz$ states.
Therefore, the ratio of branching fractions and the \CP asymmetry in the \SUP mode are measured separately in the full phase space and in a restricted phase-space region 
which involves $\Lb \to \D X$ decays, where an enhanced sensitivity to $\gamma$ is expected.

\section{Detector and simulation}
\label{sec:Detector}

The \lhcb detector~\cite{LHCb-DP-2008-001,LHCb-DP-2014-002} is a single-arm forward
spectrometer covering the \mbox{pseudorapidity} range $2<\eta <5$,
designed for the study of particles containing \bquark or \cquark
quarks. The detector includes a high-precision tracking system
consisting of a silicon-strip vertex detector surrounding the $pp$
interaction region~\cite{LHCb-DP-2014-001}, a large-area silicon-strip detector located
upstream of a dipole magnet with a bending power of about
$4{\mathrm{\,Tm}}$, and three stations of silicon-strip detectors and straw
drift tubes~\cite{LHCb-DP-2013-003,LHCb-DP-2017-001}
placed downstream of the magnet.
The tracking system provides a measurement of the momentum, \ptot, of charged particles with
a relative uncertainty that varies from 0.5\% at low momentum to 1.0\% at 200\gevc.
The minimum distance of a track to a primary $pp$ collision vertex (PV), the impact parameter (IP), 
is measured with a resolution of $(15+29/\pt)\mum$,
where \pt is the component of the momentum transverse to the beam, in\,\gevc.
Different types of charged hadrons are distinguished using information
from two ring-imaging Cherenkov detectors~\cite{LHCb-DP-2012-003}. 
Hadrons are identified by a calorimeter system consisting of
scintillating-pad and preshower detectors, an electromagnetic
and a hadronic calorimeter. 
The online event selection is performed by a trigger~\cite{LHCb-DP-2012-004}, 
which consists of a hardware stage, based on information from the calorimeter, followed by a software stage, which applies a full event
reconstruction.

Simulation is required to model the effects of the detector acceptance and the
imposed selection requirements.
In the simulation, $pp$ collisions are generated using
\pythia~\cite{Sjostrand:2007gs,*Sjostrand:2006za} 
with a specific \lhcb configuration~\cite{LHCb-PROC-2010-056}.
Decays of unstable particles
are described by \evtgen~\cite{Lange:2001uf}, in which final-state
radiation is generated using \photos~\cite{davidson2015photos}.
The interaction of the generated particles with the detector, and its response,
are implemented using the \geant
toolkit~\cite{Allison:2006ve, *Agostinelli:2002hh} as described in
Ref.~\cite{LHCb-PROC-2011-006}.

The particle identification (PID) response in the simulated samples is corrected using control samples of $\Lc\to p\Km\pip$ decays in LHCb data, taking into account its correlation with the kinematic properties of each track and with the event multiplicity. To parametrise the PID response, an unbinned method is employed, where the probability density functions (PDFs) are modelled using kernel density estimation~\cite{Poluektov:2014rxa}.

\section{Reconstruction and selection of candidates}
\label{sec:Selection}
Neutral $D$ meson candidates are reconstructed by combining kaon and pion candidates having opposite charge. To form \Lb candidates, the neutral $D$ meson candidates are combined with proton and kaon candidates having opposite charge. Each \Lb candidate is associated to the PV for which the value of \chisqip is minimised, where \chisqip is
the difference between the vertex \chisq of a given PV with and without the \Lb candidate included in the PV fit. The tracks forming the \Lb candidate are required to have good fit quality and to be well separated from any PV in the event. 
The invariant masses of the \Lb candidate, $M(Dp\Km)$, and of the $D$ candidate, $M(K\pi)$, are required to be in the intervals from $5200$ to $7000 \mevcc$ and $1850$ to $1880 \mevcc$, respectively. Candidates with $M(K\pi)$ in a wider mass range from $1765$ to $1965 \mevcc$ are retained to quantify the background contribution from charmless \bquark-hadron decays.
A kinematic fit is performed in which $M(K\pi)$ is constrained to
the known \Dz mass~\cite{PDG2020} and the \Lb candidate's trajectory is required to point back to the associated PV~\cite{Hulsbergen:2005pu}. To improve the resolution of the squared invariant masses $M^2(\D p)$, $M^2(\D \Km)$ and $M^2(p\Km)$, the fit is repeated when calculating these variables, with the additional constraint that the invariant mass of the $\D p \Km$ combination is equal to the known \Lb mass~\cite{PDG2020}. 
Inclusion of the mass constraint improves the resolution on the two-body masses by 50\% or more, depending upon position in the three-body phase space.

Background from $\Lb \to \Lc h^{-}$ decays, with $\Lc \to p h^{-}h^{+}$ where $h$ is a charged kaon or pion, are vetoed by requiring that the invariant mass of any $p h^{-}h^{+}$ combination differs from the known \Lc mass~\cite{PDG2020} by more than $20\mevcc$. To suppress the contribution of events from charmless $\Lb \to p h^- h^+ h^-$ decays, the decay-time significance of the $D$ meson candidates with respect to the \Lb vertex is required to be larger than 2.5. The decay-time significance of the $D$ candidate  is defined as the measured decay time divided by its uncertainty.

To suppress combinatorial background, a Boosted Decision Tree (BDT) algorithm using adaptive boosting~\cite{Breiman} is employed as implemented in the TMVA toolkit~\cite{TMVA4}. To train the BDT classifier, one for the full and another for the restricted phase space, a sample of simulated \lbtodpkfav decays is used as a proxy for signal and candidates in the \Lb mass sidebands with mass in the intervals from $5300$ to $5400 \mevcc$ and from $5900$ to $6000 \mevcc$ are used to represent combinatorial background. The variables that enter the BDT selection are the quality of the kinematic fit, the quality of the \Lb and $D$ vertices and their decay-time significances, PID variables, and the \pt of the final-state particles. The optimisation criterion used to determine the choice of BDT working point is the maximum expected statistical significance of the \SUP signal, $N_\text{sig} / \sqrt{N_\text{sig} + N_\text{bkg}}$, where $N_\text{sig}$ and $N_\text{bkg}$ are the expected numbers of signal and background candidates in the mass interval from $5600$ to $5640 \mevcc$. The expected number of suppressed signal decays is determined by dividing the observed yield in the \FAV decay by the Cabibbo suppression factor defined in Eq.~\ref{eq_Rpred} using values for the CKM matrix elements taken from~\cite{PDG2016}. In 0.1\% of events  multiple \Lb candidates are reconstructed. All the candidates are retained.

\section{Determination of signal yields}
\label{sec:Fit}

\begin{figure}[tb]
	\begin{center}
		\includegraphics[width=0.49\linewidth]{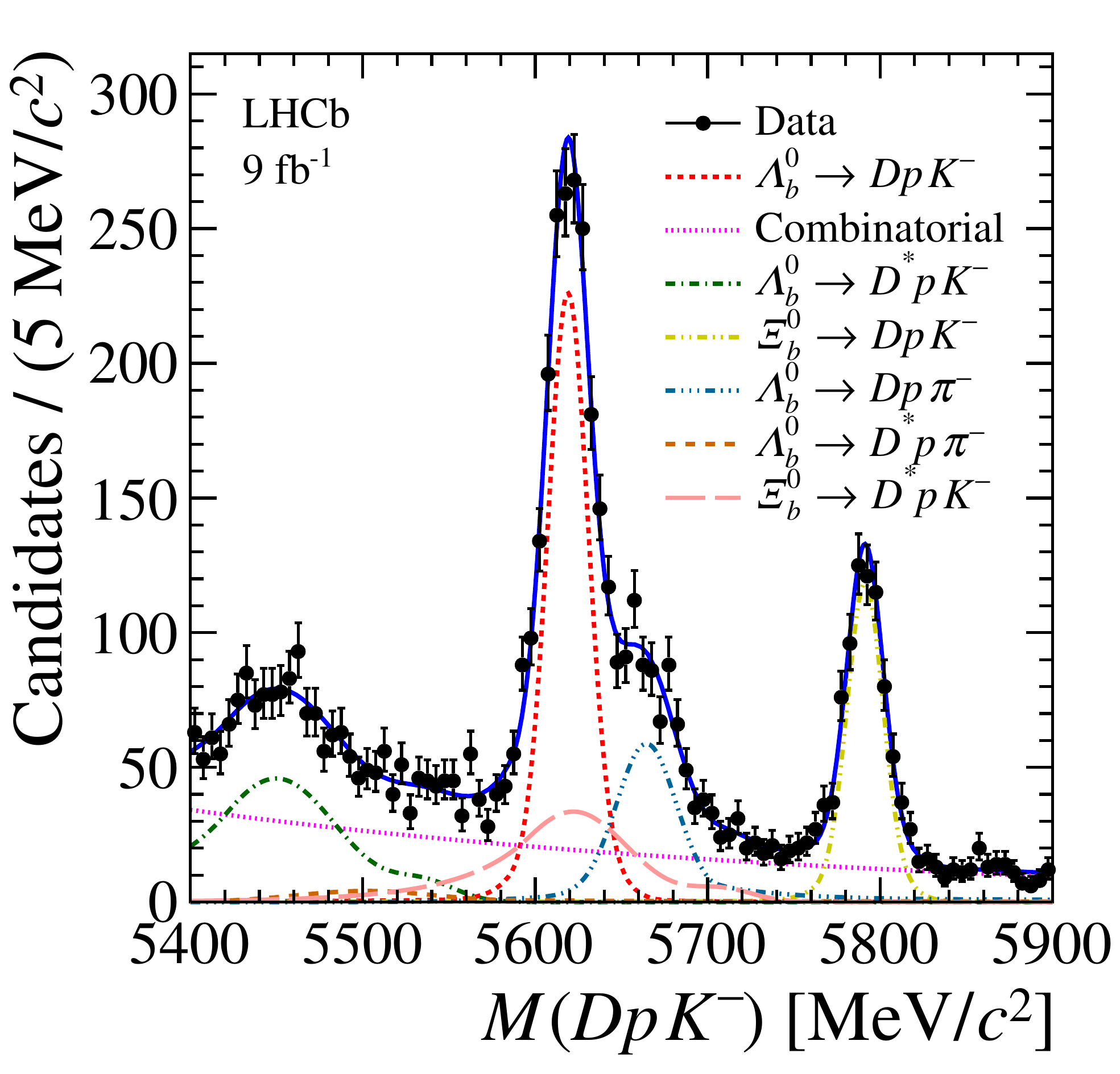}
		\includegraphics[width=0.49\linewidth]{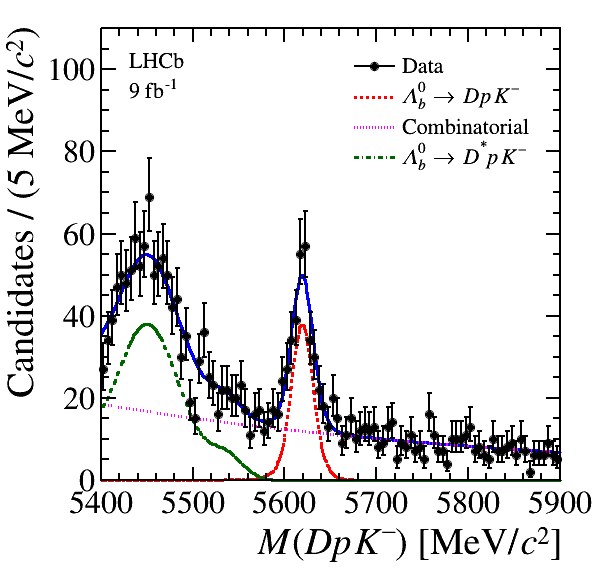}
	\end{center}
	\caption{Distributions of the invariant mass for selected (left) \lbtodpkfav and (right) \lbtodpksup candidates in the full phase space (black points) corresponding to the favoured and suppressed decays, respectively. The total fit model, as described in the text, is indicated by the solid blue line, and individual components are indicated.}
	\label{fig:fitdata}
\end{figure}

The mass distributions for the $Dp\Km$ candidates in the \FAV and \SUP data samples in the full phase space are shown in Fig.~\ref{fig:fitdata}.
The number of signal candidates is obtained by an extended unbinned maximum-likelihood fit to the $M(Dp\Km)$ mass distributions using the \textsc{RooFit} package~\cite{Verkerke:2003ir}. The \FAV and \SUP samples are fitted simultaneously. The PDF used for the \FAV sample is made of two components to model $\Lb \to \D \proton \Km$ and $\Xibz \to \D \proton \Km$ signals, a background from misidentified $\Lb\to \D p \pim$ decays, and a partially reconstructed background from $\Lb \to \Dstar \proton \Km$ decays where $\Dstarz \to \Dz \gamma$ or $\Dstarz \to \Dz \piz$ and the $\gamma$ or $\piz$ particle is not reconstructed. Additional components are required to describe the background due to partially reconstructed $\Lb \to \Dstar \proton \pim$ decays, with $\Dstarz \to \Dz \gamma$ or $\Dstarz \to \Dz \piz$ and having the pion from the \Lb decay misidentified as a kaon, partially reconstructed $\Xibz \to \Dstar \proton \Km$ decays, which peak in the signal region, and from combinatorial background. The PDF used to fit the \SUP sample is the same, except that it does not include contributions from the $\Xibz \to D^{(*)}\proton\Km$ and $\Lb \to D^{(*)}\proton\pim$ decays, which are expected to be negligible.

The parametrised shape of each component is taken from a fit to simulated decays after all selections are applied. The signal and partially reconstructed background are each modelled by the sum of two Crystal Ball (CB) functions~\cite{Skwarnicki:1986xj}, where the parameters governing the shape of the tails are fixed to their values in fits to simulated samples. The background from misidentified $\Lb\to \D p \pim$ decays is parametrised by the sum of a Gaussian and a CB function. In the restricted phase-space region, $M^2(p\Km)<5 \gevgevcccc$,
the same functional forms are used, except that the partially reconstructed background is parametrised by the sum of a bifurcated Gaussian and a CB function. The combinatorial background is described by an exponential function. The slope of the combinatorial background is allowed to vary independently in the \FAV and \SUP samples. The widths of each peaking component are multiplied by a common free parameter in order to account for the difference between the invariant-mass resolution observed in data and simulation. The mass of the \Lb baryon is a free parameter, while the mass difference between \Xibz and \Lb baryons is fixed to its known value~\cite{PDG2020}. The yields of each component are allowed to vary independently in the \FAV and \SUP samples.

\begin{figure}[tb]
	\begin{center}
		\includegraphics[width=0.49\linewidth]{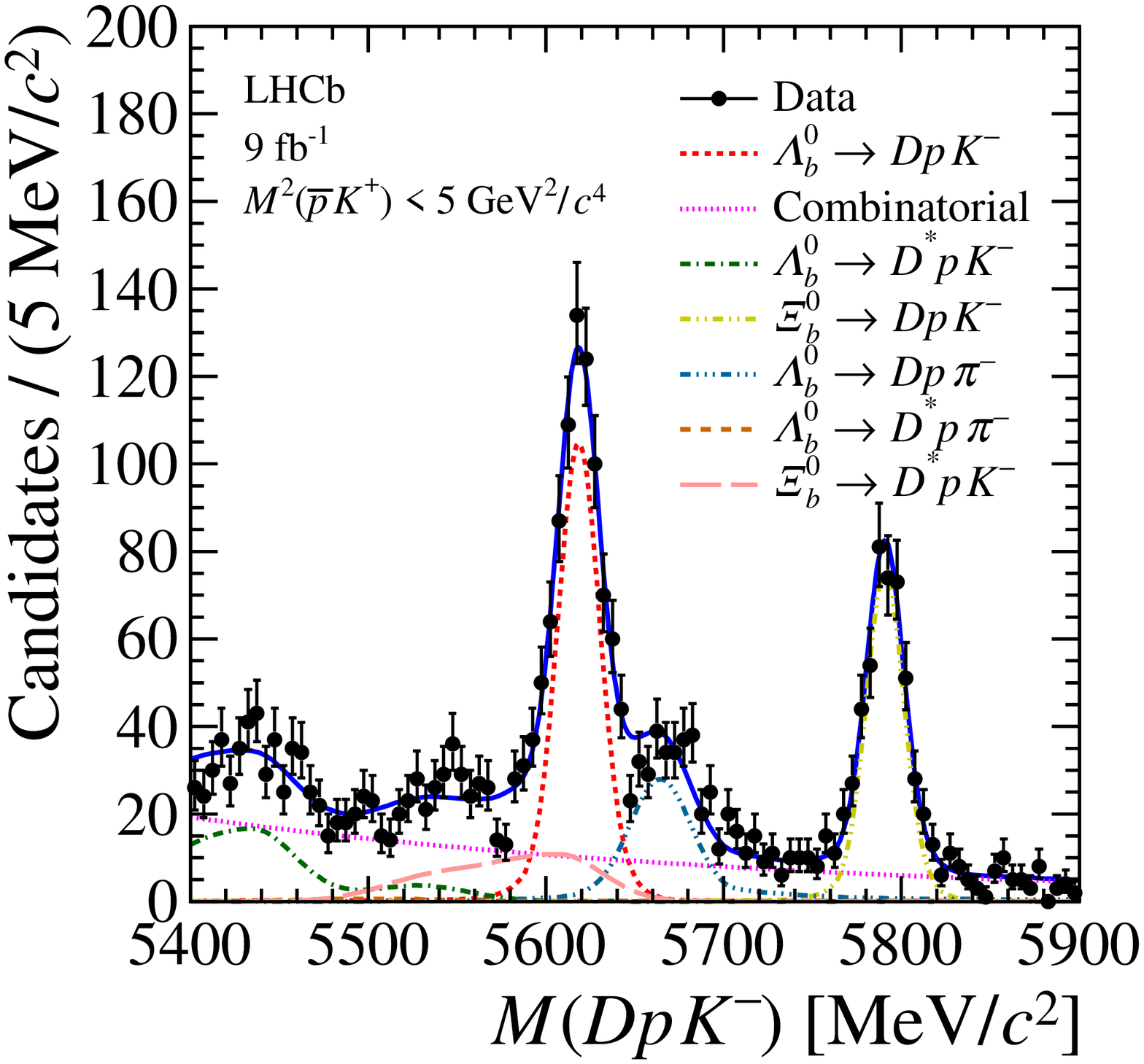}
		\includegraphics[width=0.49\linewidth]{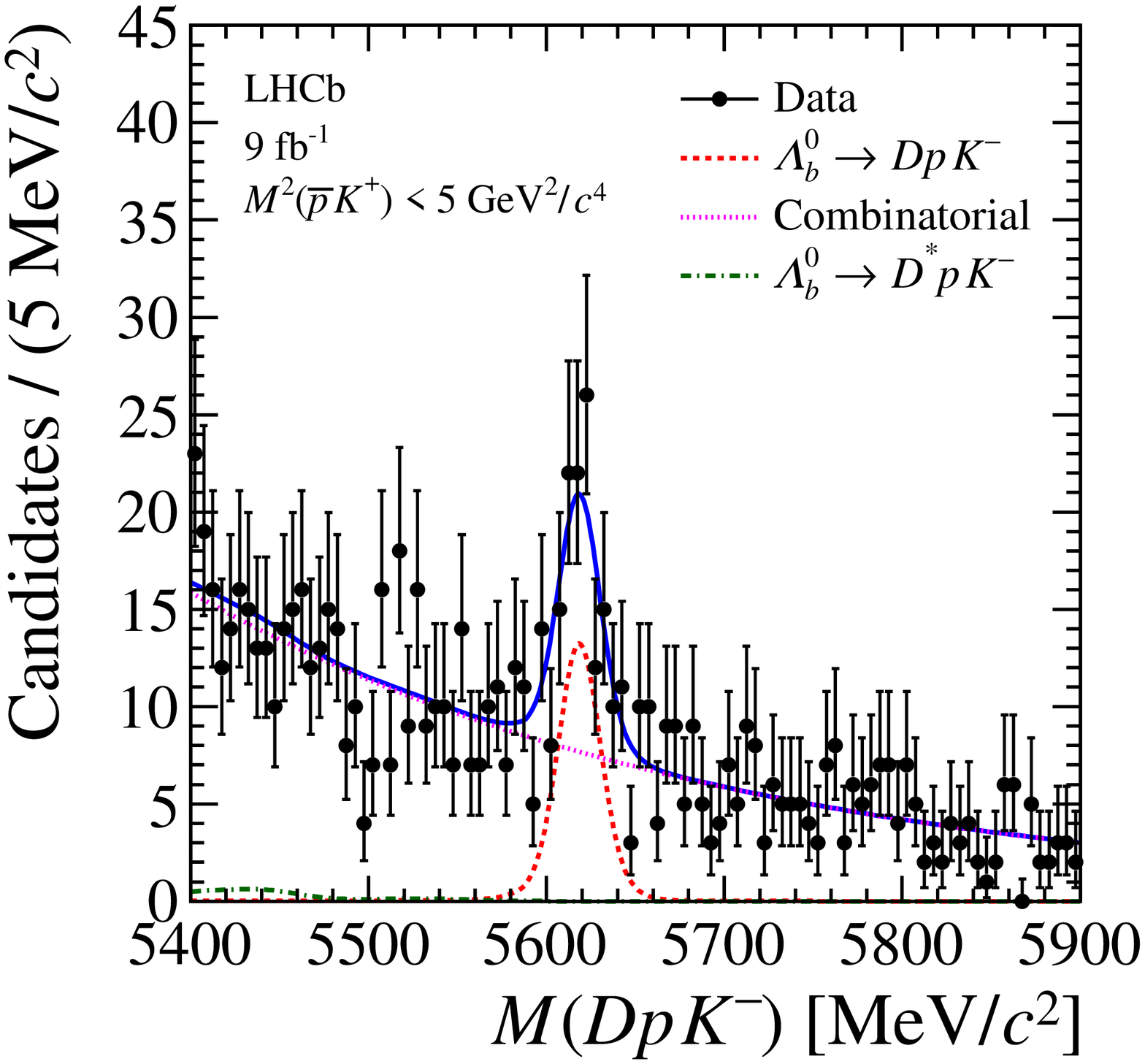}
	\end{center}
	\caption{Distributions of the invariant mass for (left) \lbtodpkfav and (right) \lbtodpksup candidates,  corresponding to the favoured and suppressed decays, respectively, in the restricted phase-space region $M^2(p\Km)<5 \gevgevcccc$. The fit, as described in the text, is overlaid.}
	\label{fig:fitdata_lowmpk}
\end{figure}

\begin{figure}[tb]
	\begin{center}
		\includegraphics[width=0.49\linewidth]{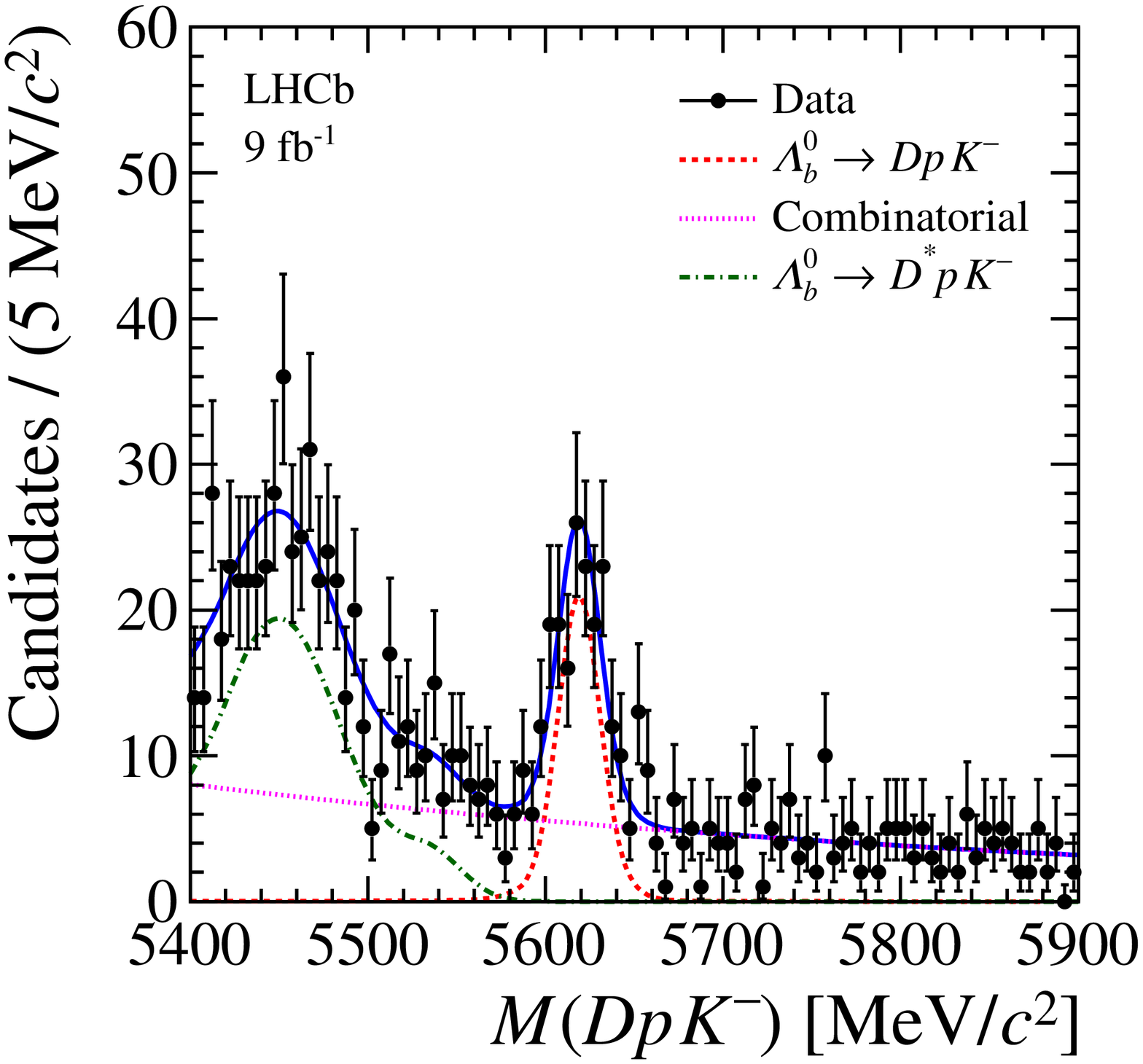}
		\includegraphics[width=0.49\linewidth]{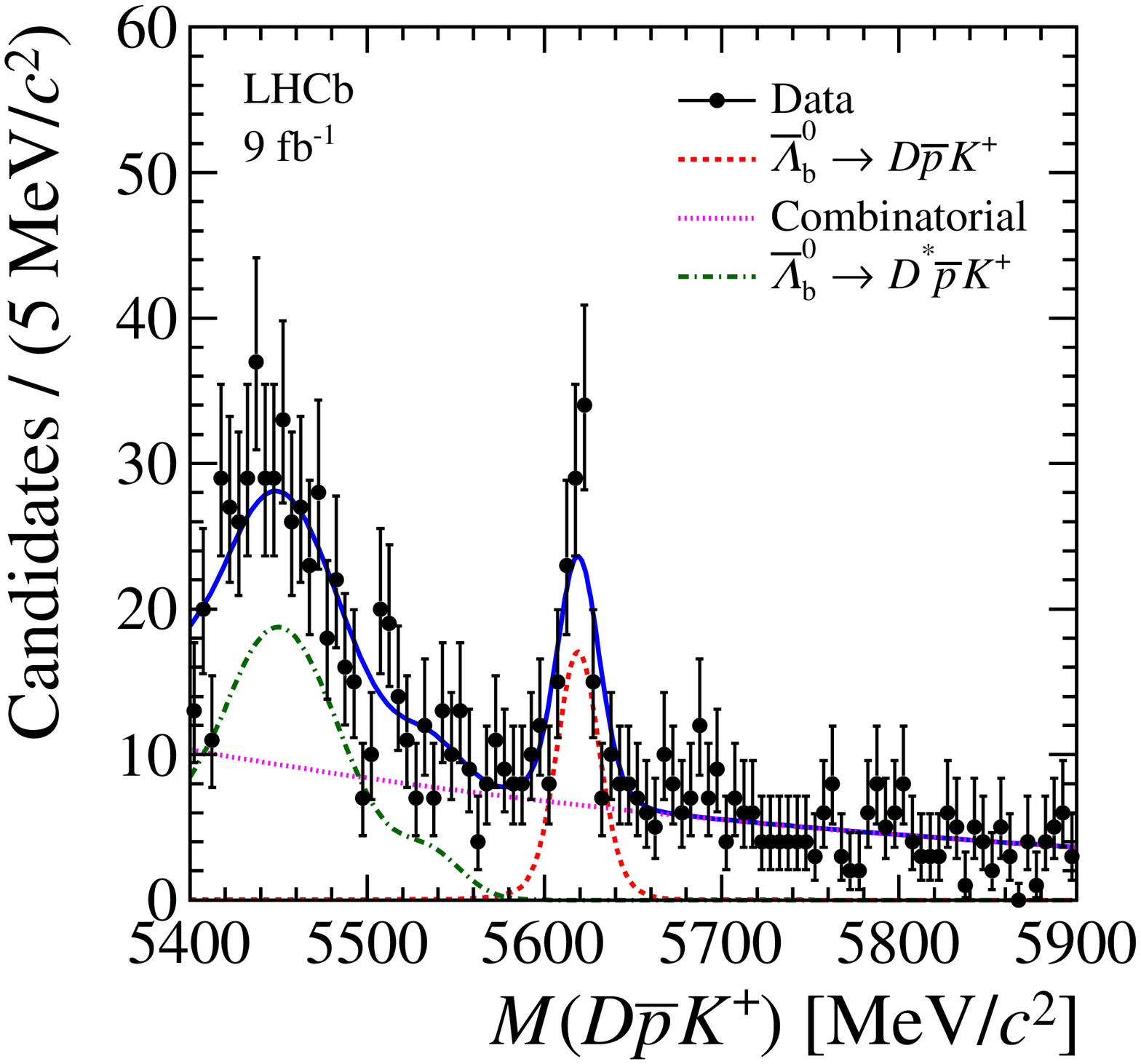}
	\end{center}
	\caption{Distributions of the invariant mass for (left) \lbtodpksup and (right) \lbtodpksupbar candidates in the full phase space (black points), corresponding to separation of the suppressed decay sample by \Lb flavour. The fit, as described in the text, is indicated by the solid blue line, and individual components are indicated.}
	\label{fig:fitdata_sup_lbbar}
\end{figure}

\begin{figure}[tb]
	\begin{center}
		\includegraphics[width=0.49\linewidth]{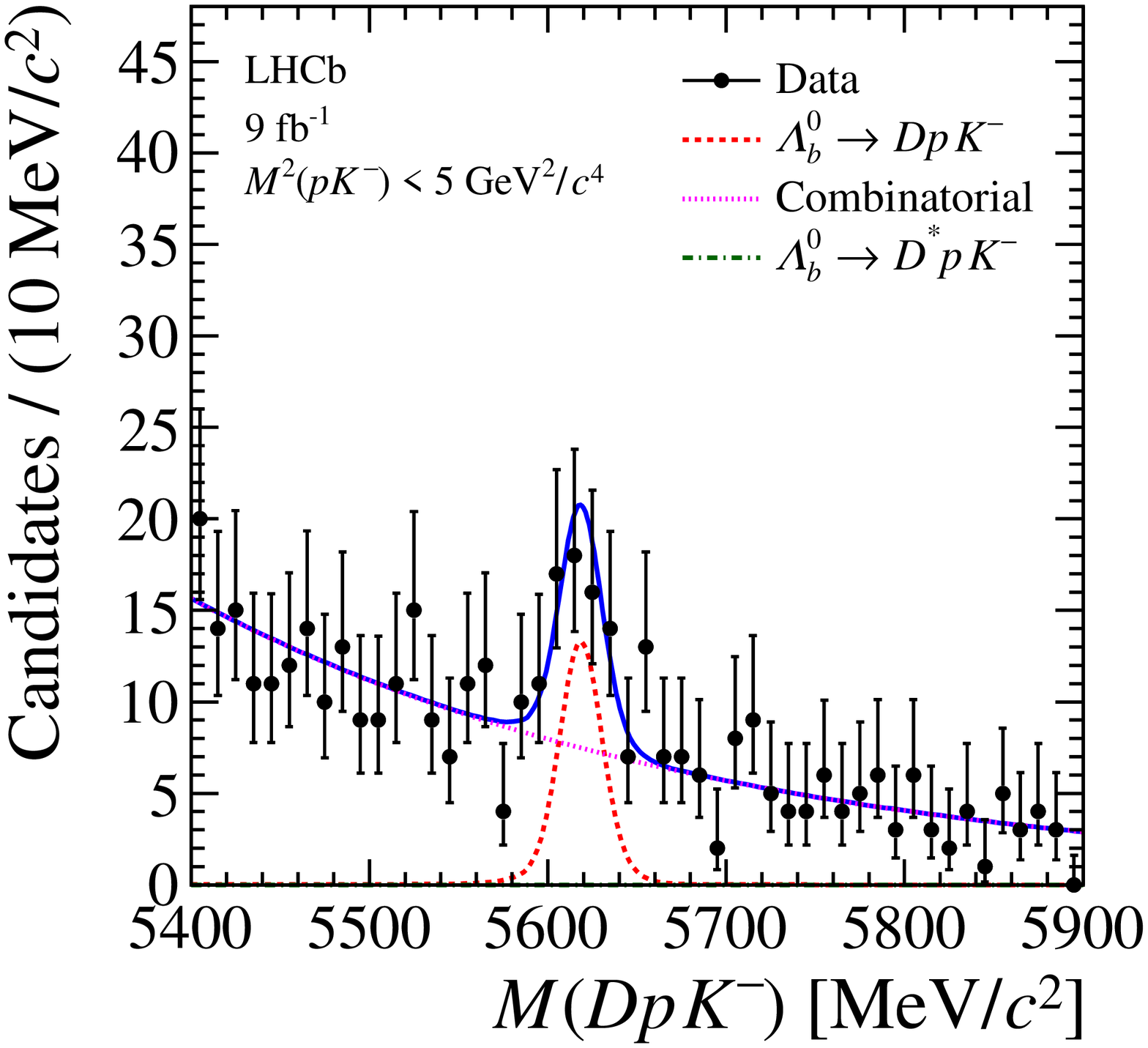}
		\includegraphics[width=0.49\linewidth]{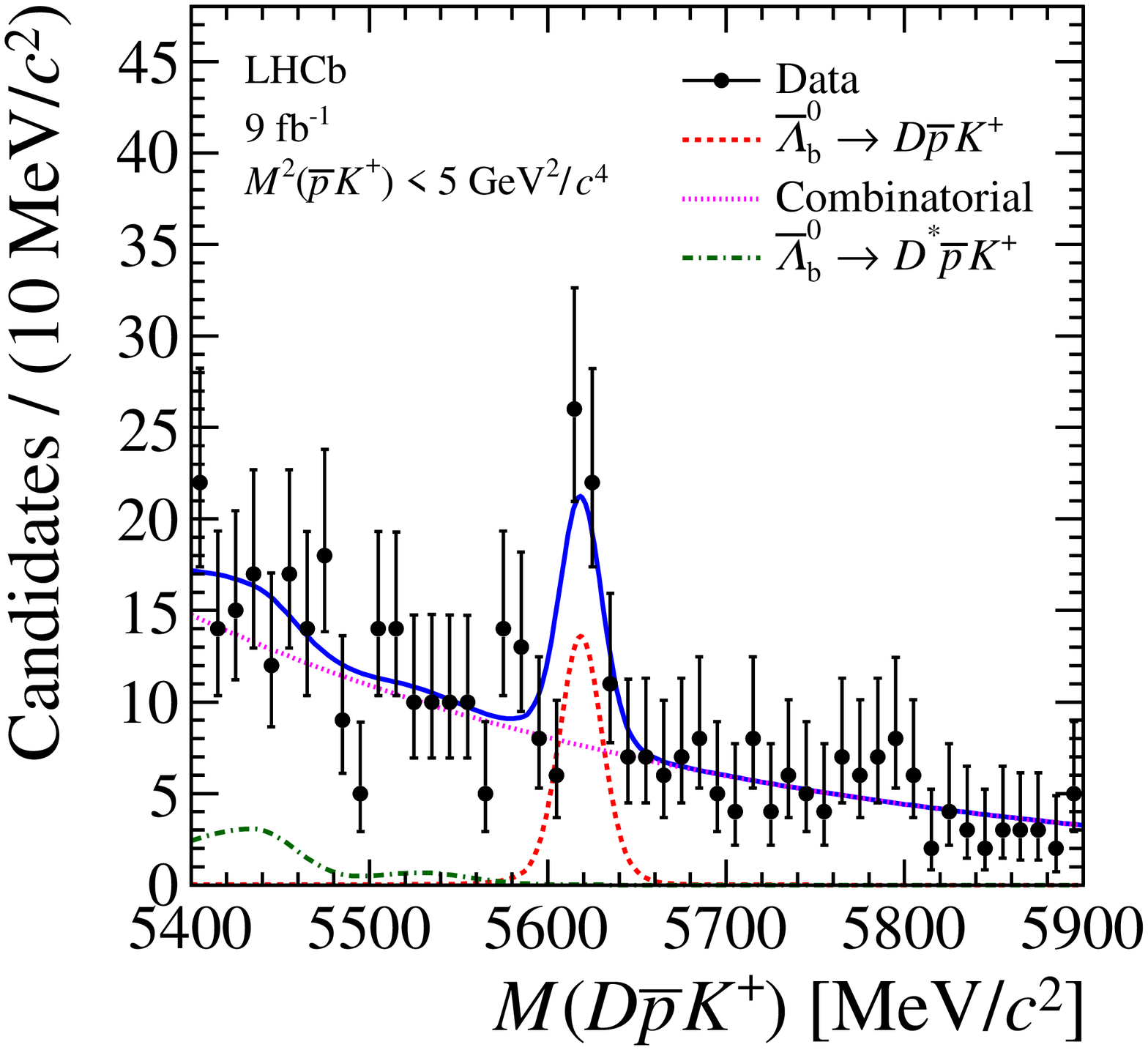}
	\end{center}
	\caption{Distributions of the invariant mass for (left) \lbtodpksup and (right) \lbtodpksupbar candidates in the restricted phase-space region (black points), where the separation is made according to the \Lb flavour. The fit, as described in the text, is indicated by the solid blue line, and individual components are indicated.}
	\label{fig:fitdata_lowmpk_sup_lbbar}
\end{figure}

The projection of the fit to the invariant-mass distribution $M(Dp\Km)$ in the \FAV and \SUP data samples in the full phase space is shown in Fig.~\ref{fig:fitdata}.
The \Lb yields are given in Table~\ref{tab:signalYieldsFromFit}. 
Figure~\ref{fig:fitdata_lowmpk} shows the invariant-mass distribution $M(Dp\Km)$ in the \FAV and \SUP data samples in the restricted phase-space region $M^2(p\Km)<5 \gevgevcccc$ with the fit projections overlaid. The signal yields obtained from this fit are given in Table~\ref{tab:signalYieldsFromFit}. 
The invariant-mass distributions $M(Dp\Km)$ and $M(D\antiproton\Kp)$, overlaid with the fit projections, used to calculate the \CP asymmetry of the \SUP decay in the full phase space and in the restricted phase-space region are shown in Figs.~\ref{fig:fitdata_sup_lbbar} and \ref{fig:fitdata_lowmpk_sup_lbbar}, respectively.

\begin{table}[htbp]
    \centering
    \caption{Signal \Lb yields obtained from fits to the invariant-mass distributions, $M(Dp\Km)$, for the favoured and suppressed decay samples in the two phase-space regions.}    
    \begin{tabular}{l r @{ $\pm$ } l  r @{ $\pm$ } l}
\hline\hline
Phase-space region & \multicolumn{2}{c}{\lbtodpkfav} & \multicolumn{2}{c}{\lbtodpksup}\\
\hline
Full & 1437 & 92 & 241 & 22\\
Restricted & 664 & 36 & 84 & 14\\
\hline\hline
    \end{tabular}
    \label{tab:signalYieldsFromFit}
\end{table}

\begin{figure}[tb]
	\begin{center}
		\includegraphics[width=0.49\linewidth]{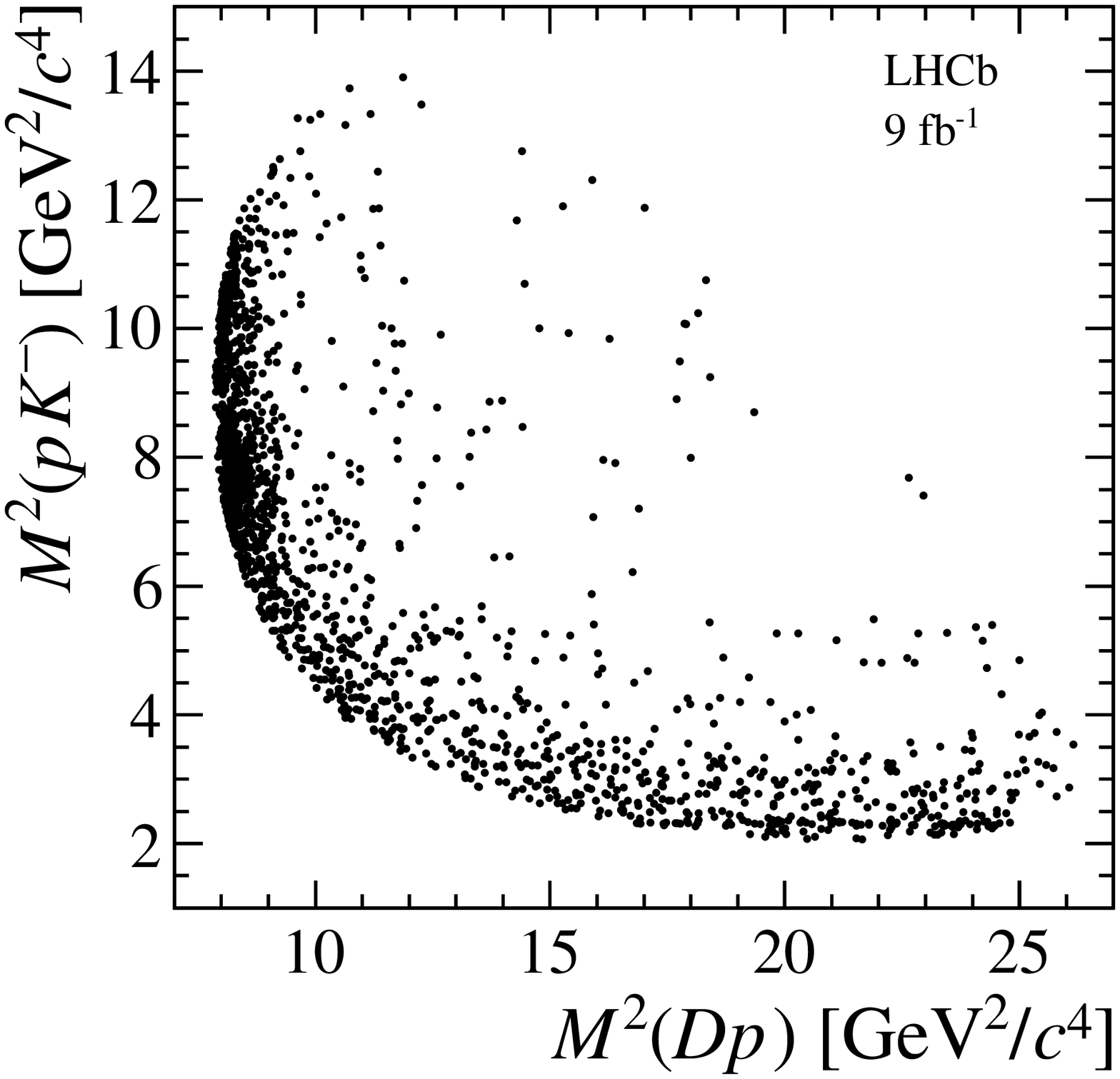}
		\includegraphics[width=0.49\linewidth]{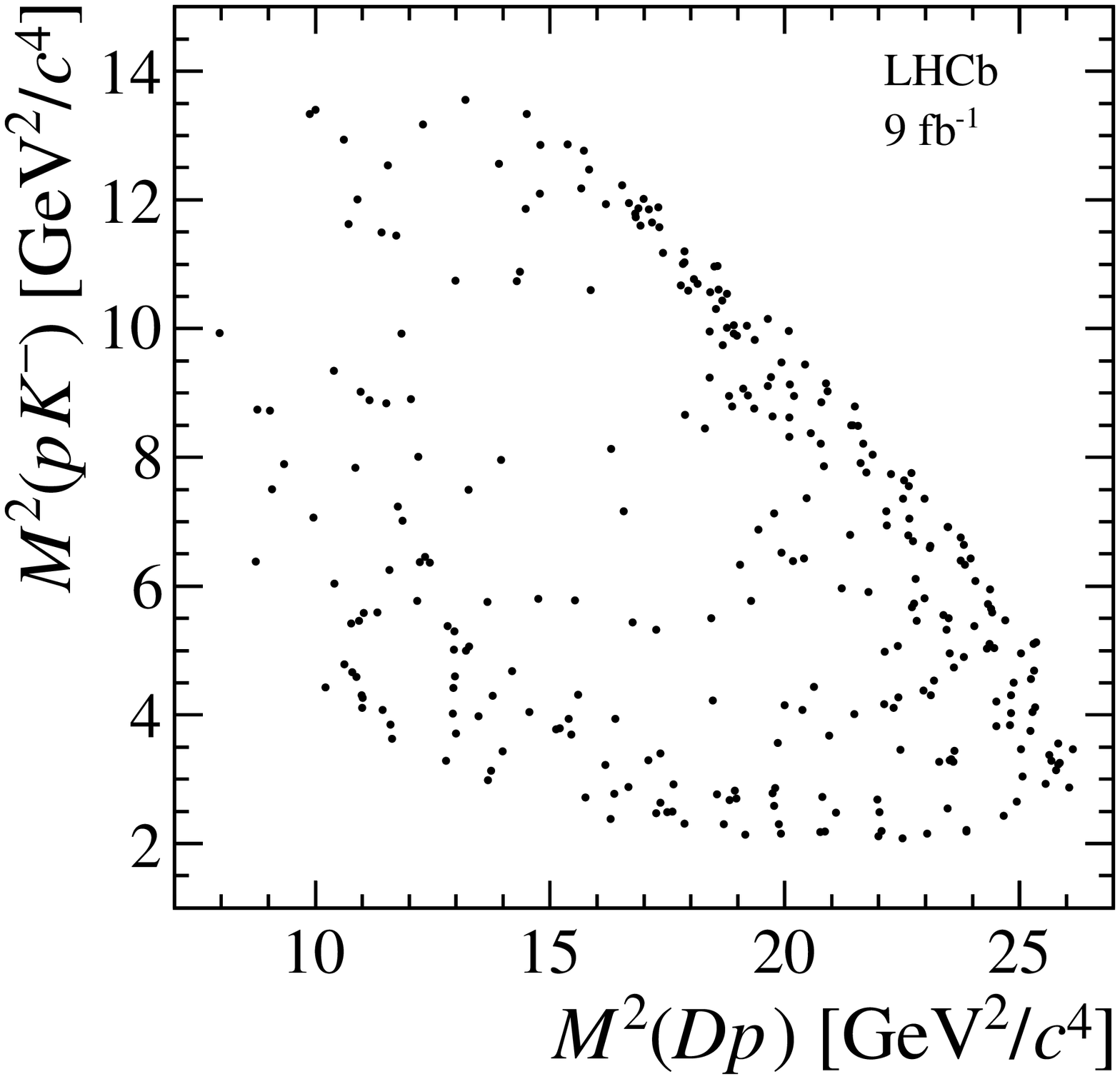}
	\end{center}
	\caption{Phase space of the (left) \lbtodpkfav and (right) \lbtodpksup candidates having a mass between 5600 and 5640\mevcc. No background subtraction is applied.}
	\label{fig:dp}
\end{figure}

\begin{figure}[tb]
	\begin{center}
		\includegraphics[width=0.325\linewidth]{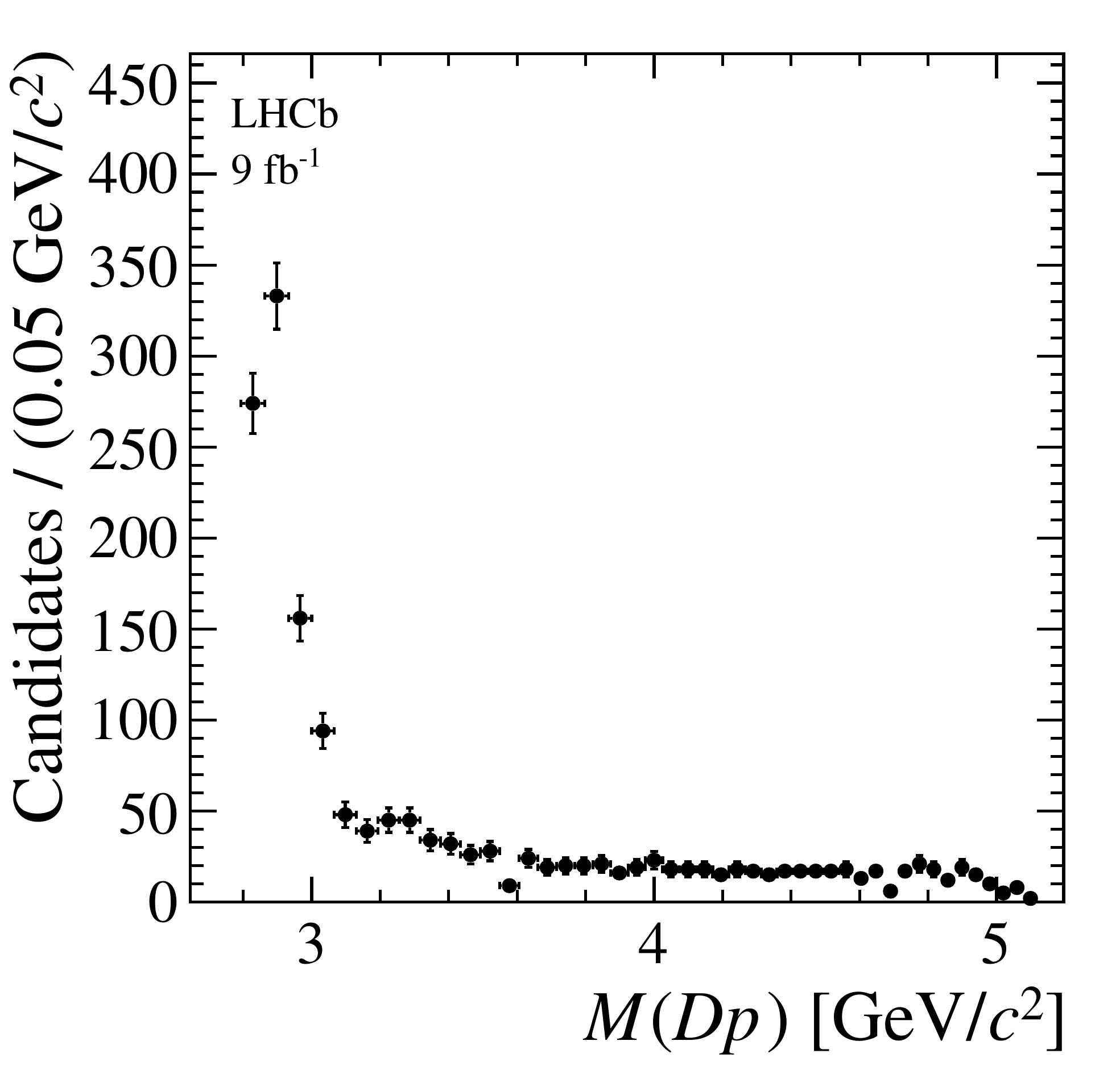}
		\includegraphics[width=0.325\linewidth]{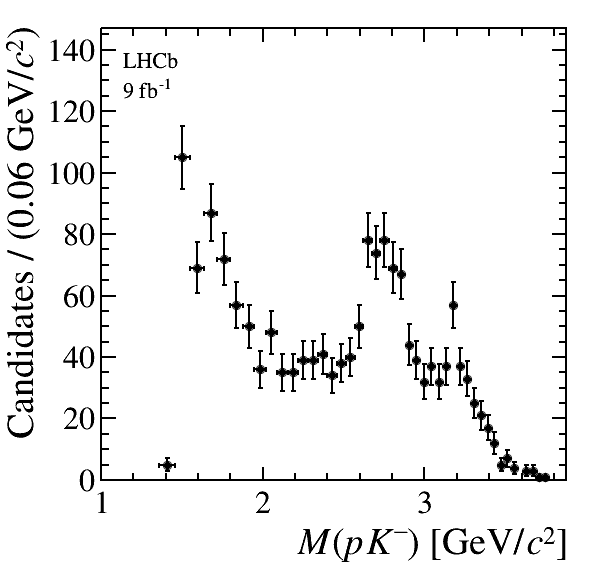}
		\includegraphics[width=0.325\linewidth]{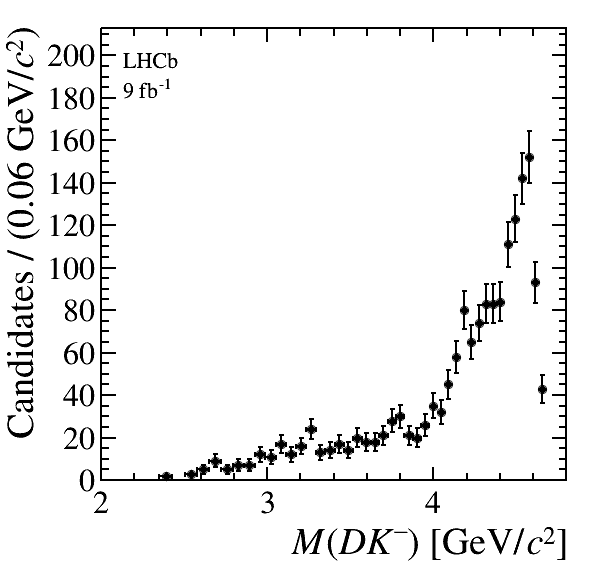}
	\end{center}
	\caption{Invariant-mass projections of the \Lb phase space in (left) $M(\D\proton)$, (middle) $M(\proton\Km$), and (right) $M(\D\Km$) for the \lbtodpkfav candidates.}
	\label{fig:dp_proj_fav}
\end{figure}

\begin{figure}[tb]
	\centering
	\includegraphics[width=0.325\linewidth]{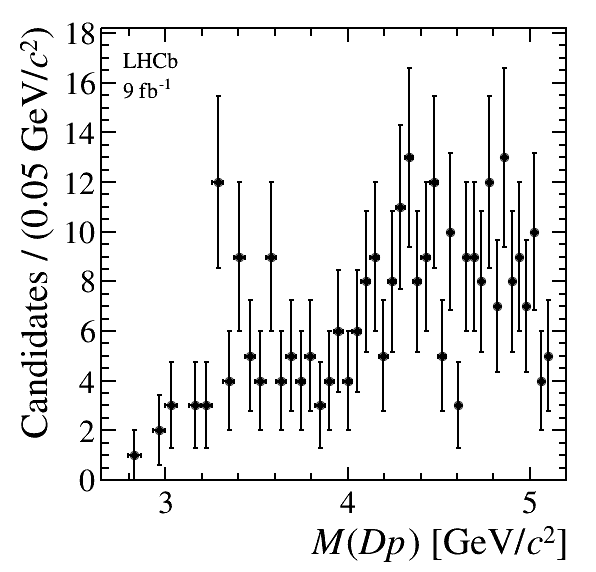}
	\includegraphics[width=0.325\linewidth]{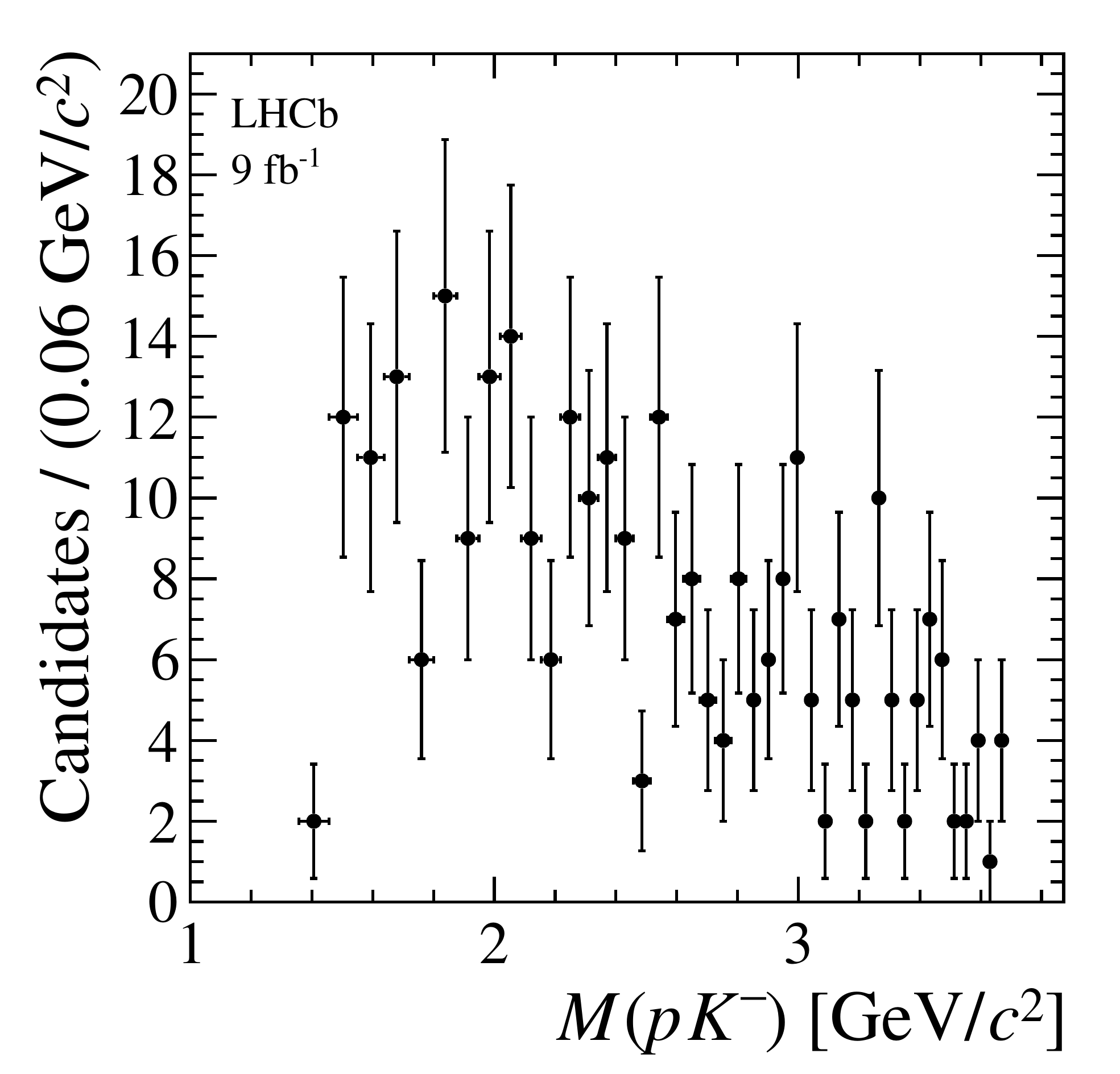}
	\includegraphics[width=0.325\linewidth]{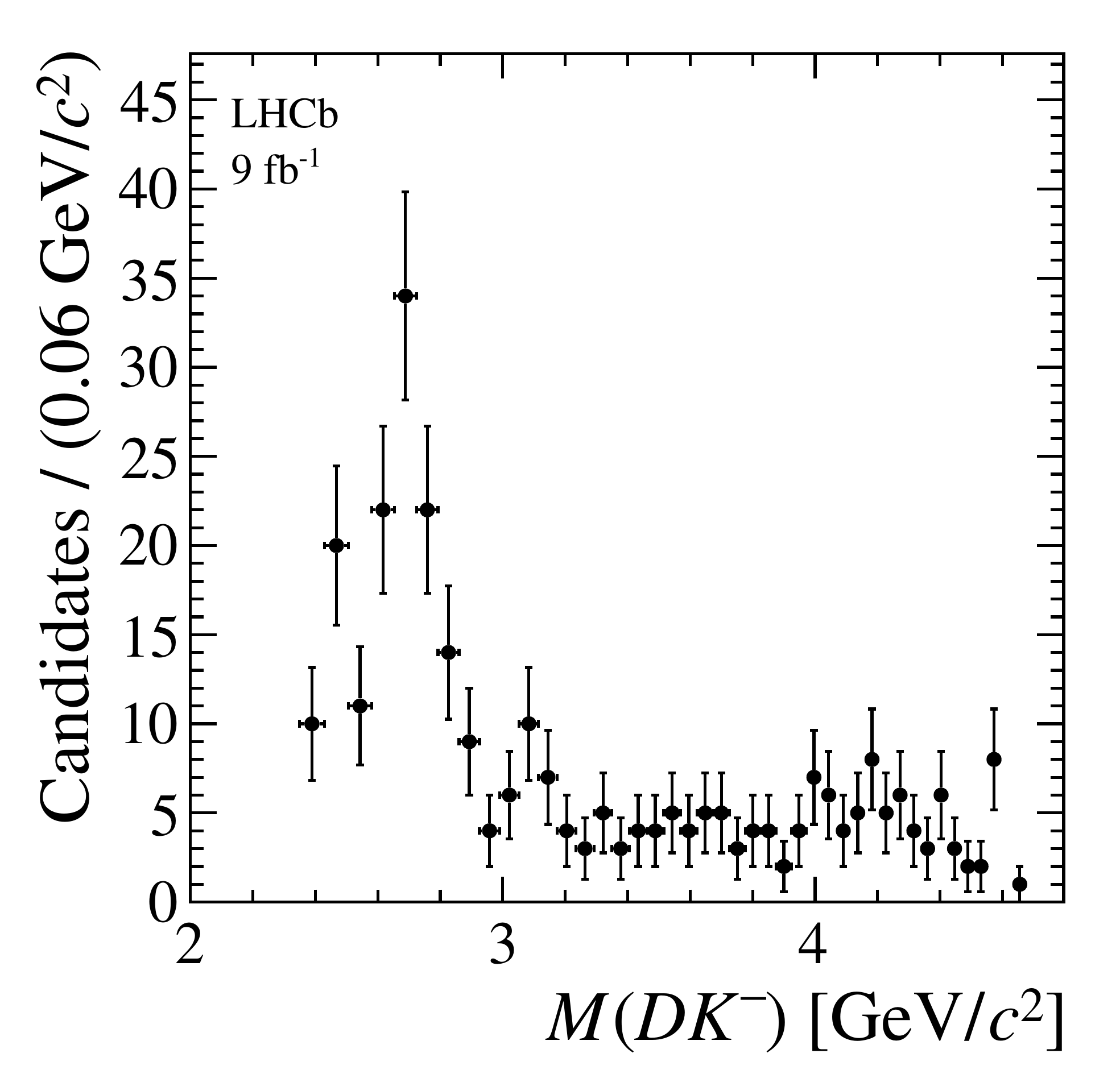}
	\caption{Invariant-mass projections of the \Lb phase space in (left) $M(\D\proton)$, (middle) $M(\proton\Km$), and (right) $M(\D\Km$) for the \lbtodpksup candidates.}
	\label{fig:dp_proj_sup}
\end{figure}

The resonant structure of the \FAV and \SUP decays can be illuminated by considering projections of the \Lb phase space.
Figure~\ref{fig:dp} shows the $M^2(\proton\Km)$ versus $M^2(\D\proton)$ distributions of \FAV and \SUP candidates in the signal region $5600 < M(\D\proton\Km) < 5640 \mevcc$. The signal purity in this region is 76\% and 72\% for the favoured and suppressed modes, respectively. Despite the smaller combinatorial background, the signal purity for the \FAV decay is comparable to that for the \SUP decay due to the presence of the background from $\Xibz \to \Dstar \proton \Km$ and $\Lb\to \D p \pim$ decays in the signal region. Figures~\ref{fig:dp_proj_fav} and \ref{fig:dp_proj_sup} show invariant-mass projections onto $M(\proton\Km)$, $M(\D\proton)$ and $M(\D\Km)$ for selected candidates. The dominant resonant amplitudes in the \FAV mode could generate structures in the $M(\D\proton)$ and $M(\proton\Km)$ distributions, corresponding to states having $udc$ and $uds$ quark content, respectively. The $M(\D\proton)$ distribution shows an increased density of events in the low-$M(\D\proton)$ region with a contribution from \mbox{$\Lambdares_c(2860)^{+}\to\Dz\proton$} decays. The distribution of $M(\proton\Km)$ contains a contribution from the $\Lz(1520)$ baryon at low-$M(\proton\Km)$ and an enhancement at \mbox{$2.5 \lesssim M(\proton\Km) \lesssim 3.5 \gevcc$}, which is the reflection of the $\Lambdares_c(2860)^{+}$ resonance seen in the $M(\D\proton)$ distribution. Different resonant structure is anticipated in the \SUP sample given the contributions from, and interference between, the $\Lb\to\Dz\proton\Km$ and $\Lb\to\Dzb\proton\Km$ amplitudes. The $M(\D\Km)$ distribution in the \SUP sample shows an increased density of events in the low-mass region with a contribution from resonances decaying to $D\Km$ such as the $D_{s1}^{*}(2700)^{\pm}$.

\section{Calculation of branching fraction ratio and \CP asymmetry}
\label{sec:Eff}

\begin{figure}[tb]
	\centering
	\includegraphics[width=0.7\linewidth]{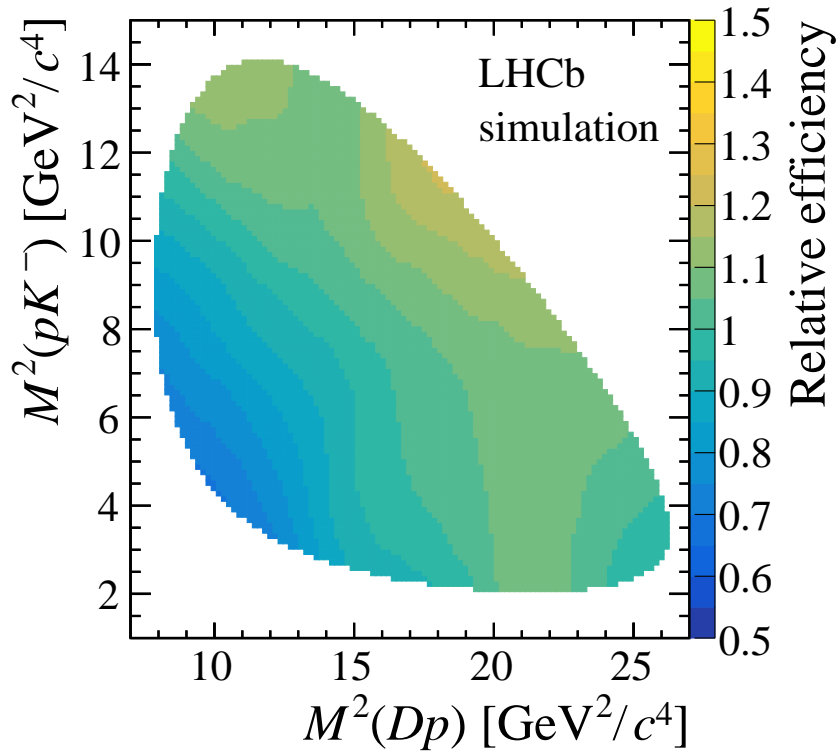}
	\caption{Relative efficiency correction as a function of position within the \Lb phase space, obtained from simulated samples.}
	\label{fig_eff}
\end{figure}

The ratio of branching fractions of the \FAV and \SUP decays and the \CP asymmetry of the \SUP decay are calculated from the ratio of yields of the corresponding decays after applying efficiency correction factors as
\begin{align}
\label{eq:RA}
R &= \frac{\sum_i w^i_{\text{FAV}}/\epsilon^i}{\sum_i w^i_{\text{SUP}}/\epsilon^i},\\
A &= \frac{\sum_i w^i_{\text{SUP,$\Lb$}}/\epsilon^i-\sum_i w^i_{\text{SUP,$\Lbbar$}}/\epsilon^i}{\sum_i w^i_{\text{SUP,$\Lb$}}/\epsilon^i+\sum_i w^i_{\text{SUP,$\Lbbar$}}/\epsilon^i}
\end{align}
where the sum is over the selected candidates. Here $w^i_{\text{FAV}}$ and $w^i_{\text{SUP}}$ are the weights obtained using the \sPlot technique~\cite{Pivk:2004ty}  for background subtraction of the \FAV or \SUP samples, respectively, with $M(Dp\Km)$ as the discriminating variable. The subscripts \Lb and \Lbbar label the samples split by flavour and $\epsilon^i$ are the relative efficiencies. 
The efficiency corrections are determined as a function of the \Lb phase-space variables $M^2(D\proton)$ and $M^2(\proton\Km)$ using the simulated \mbox{$\Lb\to [\Km\pip]_{D} p \Km$} sample and parameterised by a kernel density estimation technique~\cite{Poluektov:2014rxa}. Across the phase space, the relative efficiencies vary from $0.7$ to $1.2$, as shown in Figure~\ref{fig_eff}.

The measured values of $R$ and $A$ in the full phase space with their statistical and systematic uncertainties are
\begin{align*}
R &= \Rmeas,\\
A &= \Ameas,
\end{align*}
and in the restricted phase-space region $M^2(p\Km)<5 \gevgevcccc$,
\begin{equation*}
\begin{split}
R &= \Rlowmpkmeas,\\
A &= \Alowmpkmeas.
\end{split}
\end{equation*}
The data samples in the full and restricted phase-space region partially overlap and the statistical uncertainties on the ratios and asymmetries measured in these two regions are correlated. A positive correlation between the statistical uncertainties of $\rho = 0.33$ is estimated, given the number of events in the two samples and their overlap.

\section{Systematic uncertainties}
\label{sec:Sys}

The systematic uncertainties on the ratio of branching fractions of the \FAV and \SUP decays and the \CP asymmetry of the \SUP decay are listed in Table~\ref{tab:RA}. For each variation, the determination of $R$ and $A$ is performed, and the difference with respect to the nominal result is taken as  systematic uncertainty. Where multiple variations are considered, the largest negative and positive deviations are taken. For the measurement of $R$ and $A$, many systematic effects cancel in the ratios. The total systematic uncertainties are obtained by summing all the contributions in quadrature.

\begin{table}[tb]
	\caption{Absolute uncertainties on the ratio of branching fractions, $R$, and \CP asymmetry, $A$, related to the sources of systematic uncertainty studied. The statistical and total systematic uncertainties are also shown.}
	\renewcommand{\arraystretch}{1.3}
	\begin{center}
\begin{tabular}{lll}
\hline\hline
	& \multicolumn{1}{c}{$R$} & \multicolumn{1}{c}{$A$} \\
	\hline
	Systematic uncertainties &  &  \\
\hline	
	$~~~$fit model & ${}^{+0.37}_{-0.15}$ & ${}_{-0.011}^{+0.000}$ \\
	$~~~$efficiency corrections & ${}^{+0.21}_{-0.24}$ & ${}^{+0.010}_{-0.008}$ \\
	$~~~$PID efficiency & ${}^{+0.08}_{-0.16}$ & ${}^{+0.001}_{-0.002}$ \\
	$~~~$hardware trigger efficiency &$\pm 0.03$ &$\pm 0.001$  \\	
	$~~~$charmless background & $\pm {0.08}$ & \\
	$~~~$double misid. background & $\pm {0.005}$ &  \\
	$~~~$single misid. background & $\pm {0.001}$ &  \\
	$~~~$\Lb production asymmetry &  & $\pm {0.015}$ \\
	$~~~$\proton detection asymmetry &  & $\pm {0.015}$ \\
	$~~~$$\pi$ detection asymmetry &  & $\pm {0.005}$ \\
	Total systematic uncertainty & ${}^{+0.43}_{-0.33}$ &   ${}^{+0.024}_{-0.026}$ \\
    \hline
	Statistical uncertainty & $\pm0.79$ & $\pm0.088$ \\
\hline\hline
\end{tabular}
	\end{center}
	\label{tab:RA}
\end{table}

To assess systematic uncertainties due to the description of signal and background contributions in the invariant mass fit model, alternative parameterisations for the \Lb mass peak, partially reconstructed background components and combinatorial background are used. The corresponding systematic uncertainties amount to $(1\text{--}5)\%$.
The efficiency corrections are impacted by the limited size of the signal simulation sample. This is assessed by varying the parameters of the kernel density estimation. The corresponding systematic uncertainties are $(1\text{--}3)\%$.

The PID response in data is obtained from calibration samples~\cite{LHCb-PUB-2016-021,LHCb-DP-2018-001}. The associated systematic uncertainty includes the kernel width variation and the uncertainty due to finite sample size of the calibration samples. The resulting uncertainty is 2\% for $R$, and less than $1\%$ for $A$.

The hardware-level trigger decision is not perfectly simulated. The impact of this mismodelling is estimated by varying the efficiency map according to a correction obtained from data control samples. The resulting systematic uncertainty is at the level of 0.5\%.

The background from charmless decays is estimated by interpolating from the $D$ mass sidebands into the $D$ mass region after all selection requirements are imposed and found to be $0.5\%$ and $1\%$ in the \FAV and \SUP samples, respectively. These values are taken as uncertainties and propagated to the measurement of $R$, resulting in an uncertainty of $1\%$ on $R$, while the corresponding uncertainty on $A$ is assumed to cancel.

Doubly misidentified background, where the kaon and pion in the favoured mode are swapped, leads to a background that peaks at the \Lb mass
in the suppressed sample.
This contribution is estimated from simulation to be $0.5\%$ in the \SUP sample, and this value is assigned as an uncertainty in $R$.
Furthermore, the impact of $\Lb \to \D\proton\Km$ decays with subsequent $\D \to \Km\Kp$ or $\D \to \pim\pip$ decays, when one of the \D decay products is misidentified, is found to be $0.1\%$.

The asymmetry in \Lb and \Lbbar production is expected to influence the measurement of $A$. 
The average values were measured to be $(1.92\pm 0.35)\%$ and $(1.09\pm 0.29)\%$ at $7$ and 8\tev centre-of-mass energies, respectively~\cite{LHCb-PAPER-2021-016}. The production asymmetry is expected to decrease still further for a centre-of-mass energy of 13\tev\cite{Braaten:2001bf,Leibovich:2014jda}. A $1.5\%$ systematic uncertainty is assigned. 
Furthermore, the detection asymmetry of the \Lb decay products is also expected to influence the measurement of $A$. The asymmetry in detecting protons versus antiprotons has been studied in detail~\cite{LHCb-PAPER-2021-016} and does not exceed $1.5\%$ for various values of proton momentum. The asymmetry of $\pip$ and $\pim$ detection was studied in Ref.~\cite{Dufour:2019zqn} and found to be less than $0.5\%$. These values are assigned as systematic uncertainties. Any influence of $\Kp$ and $\Km$ detection asymmetry is expected to cancel for the $\Kp$ meson from the \D meson and the \Km meson from the \Lb baryon.

Pseudoexperiments were used to verify that the fit used to determine the \Lb signal yields was unbiased, and no uncertainty from this source is included.

\section{Conclusion}
\label{sec:Summary}

A study of \Lb baryon decays to the $[\Kpm\pimp]_D p\Km$ final state, where $D$ indicates a superposition of \Dz and \Dzb, is reported, using a data sample corresponding to an integrated luminosity of $9\invfb$ collected with the LHCb detector. The suppressed \lbtodpksup decay is observed for the first time. The ratio of branching fractions for the \lbtodpkfav and \lbtodpksup decays, and the \CP asymmetry, are measured in the full phase space to be
\begin{equation*}
\begin{split}
R &= \Rmeas,\\
A &= \Ameas.
\end{split}
\end{equation*}
In the phase-space region $M^2(p\Km)<5 \gevgevcccc$ the ratio and \CP asymmetry are measured to be
\begin{equation*}
\begin{split}
R &= \Rlowmpkmeas,\\
A &= \Alowmpkmeas.
\end{split}
\end{equation*}

Within the uncertainties, the ratio of the \FAV and \SUP branching fractions is consistent with the estimate based on the relevant CKM matrix elements. The measured asymmetry values are consistent with zero, both in the full phase space and in the region where enhanced sensitivity to the CKM angle $\gamma$ is expected. While the present signal yields are too low to be used to extract $\gamma$, larger samples are expected to be collected by LHCb in the coming years, and the study of this
mode will contribute to the overall determination of $\gamma$.

\section*{Acknowledgements}
%
%
\noindent We express our gratitude to our colleagues in the CERN
accelerator departments for the excellent performance of the LHC. We
thank the technical and administrative staff at the LHCb
institutes.
We acknowledge support from CERN and from the national agencies:
CAPES, CNPq, FAPERJ and FINEP (Brazil); 
MOST and NSFC (China); 
CNRS/IN2P3 (France); 
BMBF, DFG and MPG (Germany); 
INFN (Italy); 
NWO (Netherlands); 
MNiSW and NCN (Poland); 
MEN/IFA (Romania); 
MSHE (Russia); 
MICINN (Spain); 
SNSF and SER (Switzerland); 
NASU (Ukraine); 
STFC (United Kingdom); 
DOE NP and NSF (USA).
We acknowledge the computing resources that are provided by CERN, IN2P3
(France), KIT and DESY (Germany), INFN (Italy), SURF (Netherlands),
PIC (Spain), GridPP (United Kingdom), RRCKI and Yandex
LLC (Russia), CSCS (Switzerland), IFIN-HH (Romania), CBPF (Brazil),
PL-GRID (Poland) and NERSC (USA).
We are indebted to the communities behind the multiple open-source
software packages on which we depend.
Individual groups or members have received support from
ARC and ARDC (Australia);
AvH Foundation (Germany);
EPLANET, Marie Sk\l{}odowska-Curie Actions and ERC (European Union);
A*MIDEX, ANR, IPhU and Labex P2IO, and R\'{e}gion Auvergne-Rh\^{o}ne-Alpes (France);
Key Research Program of Frontier Sciences of CAS, CAS PIFI, CAS CCEPP, 
Fundamental Research Funds for the Central Universities, 
and Sci. \& Tech. Program of Guangzhou (China);
RFBR, RSF and Yandex LLC (Russia);
GVA, XuntaGal and GENCAT (Spain);
the Leverhulme Trust, the Royal Society
 and UKRI (United Kingdom).




\addcontentsline{toc}{section}{References}
\bibliographystyle{LHCb}
\bibliography{main,standard,LHCb-PAPER,LHCb-CONF,LHCb-DP,LHCb-TDR}

\newpage
\centerline
{\large\bf LHCb collaboration}
\begin
{flushleft}
\small
R.~Aaij$^{32}$,
A.S.W.~Abdelmotteleb$^{56}$,
C.~Abell{\'a}n~Beteta$^{50}$,
F.J.~Abudinen~Gallego$^{56}$,
T.~Ackernley$^{60}$,
B.~Adeva$^{46}$,
M.~Adinolfi$^{54}$,
H.~Afsharnia$^{9}$,
C.~Agapopoulou$^{13}$,
C.A.~Aidala$^{87}$
% ...

{\footnotesize \it

% ...
$^{9}$Universit{\'e} Clermont Auvergne, CNRS/IN2P3, LPC, Clermont-Ferrand, France\\
% ...
$^{13}$LPNHE, Sorbonne Universit{\'e}, Paris Diderot Sorbonne Paris Cit{\'e}, CNRS/IN2P3, Paris, France\\
% ...
$^{32}$Nikhef National Institute for Subatomic Physics, Amsterdam, Netherlands\\
% ...
$^{46}$Instituto Galego de F{\'\i}sica de Altas Enerx{\'\i}as (IGFAE), Universidade de Santiago de Compostela, Santiago de Compostela, Spain\\
% ...
$^{50}$Physik-Institut, Universit{\"a}t Z{\"u}rich, Z{\"u}rich, Switzerland\\
% ...
$^{54}$H.H. Wills Physics Laboratory, University of Bristol, Bristol, United Kingdom\\
% ...
$^{56}$Department of Physics, University of Warwick, Coventry, United Kingdom\\
% ...
$^{60}$Oliver Lodge Laboratory, University of Liverpool, Liverpool, United Kingdom\\
% ...
$^{68}$Syracuse University, Syracuse, NY, United States\\
% ...
$^{87}$University of Michigan, Ann Arbor, United States, associated to $^{68}$\\
\bigskip
% ...
\medskip
}
\end{flushleft}

\end{document}